\documentstyle[sprocl,epsfig]{article}

\def\Journal#1#2#3#4{{#1} {\bf #2}, #3 (#4)}

\def\NPB{{\em Nucl. Phys.} B}
\def\PLB{{\em Phys. Lett.}  B}
\def\PRL{\em Phys. Rev. Lett.}
\def\PRD{{\em Phys. Rev.} D}
\def\ZPC{{\em Z. Phys.} C}
\def\PR{\em Phys. Rep.}

\def\pup{p^{\uparrow}}
\def\Lup{\Lambda^\uparrow} 
\def\Ldown{\Lambda^\downarrow} 
\def\qup{q^\uparrow} 
\def\bfk{\mbox{\boldmath $k$}} 
\def\bfP{\mbox{\boldmath $P$}} 
\def\bfp{\mbox{\boldmath $p$}} 
\def\be{\begin{equation}}
\def\ee{\end{equation}}
\def\bea{\begin{eqnarray}}
\def\eea{\end{eqnarray}}

\begin{document}

\title{STATUS OF SPIN PHYSICS}

\author{M. ANSELMINO}

\address{Dipartimento di Fisica Teorica, Universit\`a di Torino and \\
         INFN, Sezione di Torino, Via P. Giuria 1, 10125 Torino, Italy}

\maketitle\abstracts{
Fundamental spin physics has made striking progresses in the last years;
new ideas, experiments and data interpretations have been proposed
and keep emerging. A review of some of the most important issues
in the spin structure of nucleons is made and prospects for the future
are discussed.}

\section{Introduction}
Since the so called proton spin crisis in the parton model \cite{emc,al} -- 
a little more than a decade ago -- high energy spin physics has experienced 
an impressive blooming in the amount of proposed and performed experiments,
of new theoretical ideas, of interest in issues where spin plays a crucial
role. There is by now a general consensus on the fact that spin represents a
fundamental, quantum mechanical and relativistic, aspect of QCD field theory,
which has to be fully understood before we can claim that a good knowledge of 
the nucleon and hadron structures has been achieved.

A proof of the richness of spin physics is the difficulty in confining
the subject in half an hour talk and the written version in a limited 
numbers of pages. I'll try to choose the arguments by following the
logic line of looking into the proton structure, asking what we know about 
it. It leads us along the following path:
\begin{itemize}
\item
how and what do we know about the polarized structure functions 
$g_1(x,$ $Q^2)$ and $g_2(x,Q^2)$?
\item
how and what do we know about quark and gluon helicity distributions, 
$\Delta q(x,Q^2)$ and $\Delta g(x,Q^2)$?
\item
how and what do we know about quark and gluon orbital angular momentum,
$L_q$ and $L_g$?
\item
$\Delta q(x,Q^2)$, $\Delta g(x,Q^2)$, $L_q$ and $L_g$ are not the whole 
story: how and where do we learn about the transversity distribution 
$h_1(x,Q^2)$?
\item
Transversity, a chiral-odd function, needs other chiral-odd functions
in order to be accessible, and this leads to azimuthal asymmetries in
semi-inclusive DIS, $\ell \pup \to \ell \pi X$, or to a measurement of 
$\Lambda$ polarization, $\ell \pup \to \ell \Lup X$.
\end{itemize}

At the end, I would like to make also some comments on the ``small'' $Q^2$
transition region, where interesting results have recently been obtained 
on the Bloom-Gilman duality for $g_1$ and on the electromagnetic elastic 
proton form factors. I will not have time to comment on the unexplored 
regions that the planned $\nu$-factory experiments \cite{nf} -- neutrino 
initiated DIS with {\it polarized} targets -- might open in the not so near 
future. Lack of time and space also force me to ignore several 
other important issues.
 



\section{Polarized structure functions}

The polarized structure functions $g_1$ and $g_2$ appear in the antisymmetric
part of the hadronic tensor and are measured by combining polarized DIS
cross-sections: 
\be
{d^2\sigma^{\begin{array}{c}\hspace*{-0.1cm}\to\vspace*{-0.2cm}\\
\hspace*{-0.1cm}\Leftarrow\end{array}}\over d\Omega~dE^\prime} - 
{d^2\sigma^{\begin{array}{c}\hspace*{-0.1cm}\to\vspace*{-0.2cm}\\
\hspace*{-0.1cm}\Rightarrow\end{array}}\over
d\Omega~dE^\prime} = 
\frac{4\alpha^2 E^\prime}{Q^2 E \, M \, \nu} 
\left[(E + E^\prime\cos\theta)\,g_1 - 2Mx \, g_2\right] \,
\ee
and 
\be
{d^2 \sigma^{\to\Downarrow}\over d\Omega~dE^\prime} - 
{d^2 \sigma^{\to\Uparrow}\over d\Omega~dE^\prime} =
\frac{4 \alpha^2 E^{\prime 2}}{Q^2 E \, M \, \nu}\,\sin\theta \,\left[ g_1 + 
\frac{2E}{\nu} \, g_2\right]\,, 
\ee
which are written in terms of typical DIS variables, following 
the usual conventions for longitudinal ($\to, \Rightarrow, \Leftarrow)$ 
and transverse polarizations ($\Uparrow, \Downarrow$) \cite{ael}.

The polarized structure functions $g_1$ and $g_2$ are actually extracted 
from data on double spin asymmetries,
\be
A_{\parallel} \equiv \frac 
  {d\sigma^{\begin{array}{c}\hspace*{-0.1cm} \to \vspace*{-0.2cm}\\
\hspace*{-0.1cm}\Leftarrow\end{array}} 
-  d\sigma^{\begin{array}{c}\hspace*{-0.1cm} \to \vspace*{-0.2cm}\\ 
\hspace*{-0.1cm}\Rightarrow\end{array}}}
  {d\sigma^{\begin{array}{c}\hspace*{-0.1cm} \to \vspace*{-0.2cm}\\
\hspace*{-0.1cm}\Rightarrow\end{array}} 
+  d\sigma^{\begin{array}{c}\hspace*{-0.1cm} \to \vspace*{-0.2cm}\\
\hspace*{-0.1cm}\Leftarrow\end{array}}} 
\quad\quad\quad
A_\perp \equiv \frac {d\sigma^{\to\Downarrow} - d\sigma^{\to\Uparrow}} 
{d\sigma^{\to\Uparrow} + d\sigma ^{\to\Downarrow}} \>,
\ee
and by now a good amount of information is available on $g_1$ [see Fig. 1]
and some first information on $g_2$ \cite{g2} [see Fig. 2]. 
This information is interpreted in the framework of the QCD parton model.

\section{Extraction of \mbox{\boldmath $\Delta q$} and 
\mbox{\boldmath $\Delta g$}}

At NLO in the QCD parton model the structure function $g_1$ is given by
\be
g_1(x, Q^2) = 
\frac 12 \sum_q e_q^2 \left\{ \Delta C_q \otimes \left[ \Delta q + 
\Delta\bar q \right] + \frac {1}{N_f} \Delta C_g \otimes \Delta g \right\} 
\ee
where $\Delta q(x, Q^2)$ and $\Delta g(x, Q^2)$ are respectively 
the quark (of flavour $q$) and gluon helicity distributions; we 
have, as usual, defined the convolution
\be 
\Delta C \otimes \Delta q \equiv \int_x^1 \frac {dy}{y} \> 
\Delta C \!\! \left( \frac xy, \alpha_s \right) \> \Delta q(y, Q^2)
\ee
and the coefficients functions $\Delta C_i$ have a perturbative
expansion
\be
\Delta C_i(x, \alpha_s) = \Delta C_i^0(x) + \frac {\alpha_s(Q^2)}{2\pi} \,
\Delta C_i^{(1)}(x) + \cdots 
\ee

\begin{figure}[t]
\begin{center}
\epsfig{file=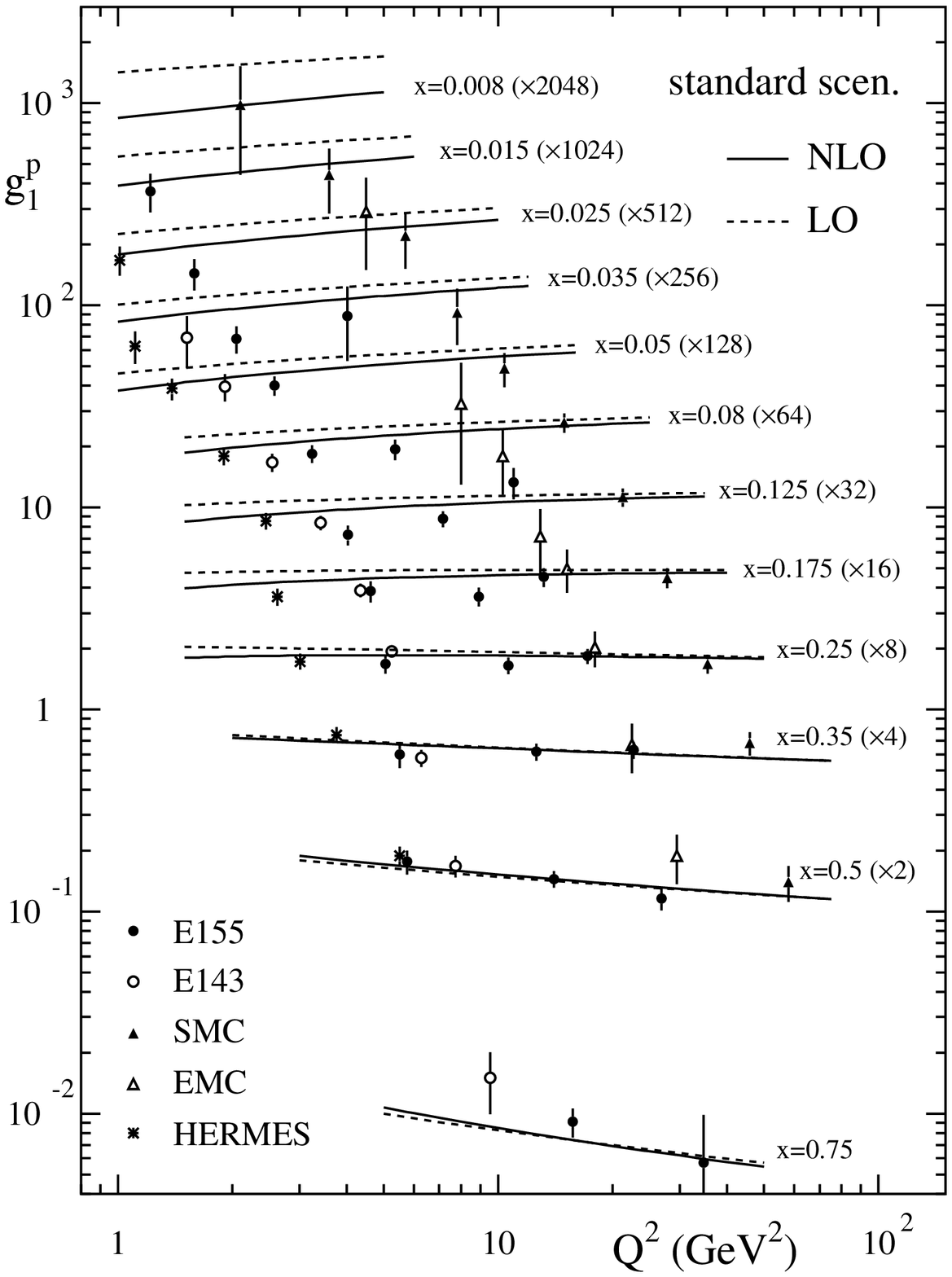,width=0.5\textwidth}\hspace*{5pt}
\epsfig{file=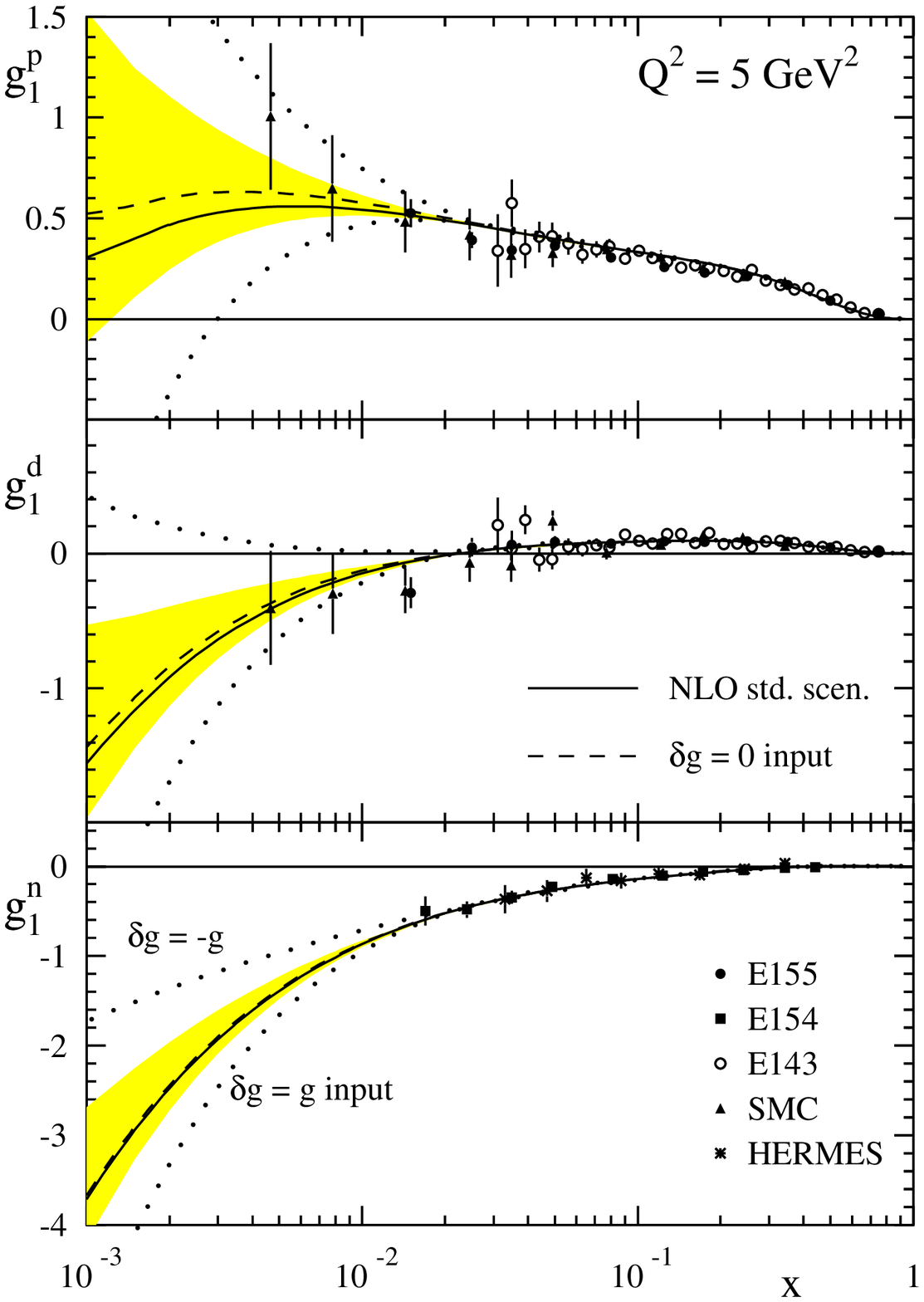,width=0.5\textwidth}
\caption{World data on $g_1$ and best fits from Ref. [8].}
\end{center}
\end{figure}

The LO terms are simply
\be
\Delta C_q^0 = \delta (1-x) \quad\quad\quad\quad \Delta C_g^0 = 0 \,,
\ee
and the NLO corrections are scheme dependent; typical choices differ in 
the amount of gluon contribution to the quark singlet distributions, while 
quark non-singlet distributions are scheme independent \cite{fj}. Finally, 
the $Q^2$ evolution of the parton densities obeys the DGLAP evolution 
equations \cite{dglap}, and,  if known at an initial scale $\mu^2$, the 
r.h.s. of Eq. (4) can be computed at any perturbative $Q^2$ value. 

By comparing data on $g_1(x, Q^2)$ with Eq. (4) one obtains information on
the quark and gluon helicity distributions; the more data one has and the 
wider the $x$ and $Q^2$ range is, the more stringent the comparison is. 
The normal procedure is that of using a simple ansatz for the unknown 
distribution functions at the initial scale $\mu^2$, with some assumptions 
regarding the sea quark densities (for example, whether $SU(3)_F$ symmetric 
or not) and some constraints from $SU(3)_F$ hyperon decay sum rules on
the first moments 
$\Delta q(1,Q^2) \equiv \int_0^1 \Delta q(x,Q^2) \> dx$ \cite{fj,vog}:
\bea
\Delta q_3 &\equiv& \Delta u(1) + \Delta \bar u(1) 
- \Delta d(1) - \Delta \bar u(1) = 1.2670 \pm 0.0035 \\
\Delta q_8 &\equiv& \Delta u(1) + \Delta \bar u(1) 
+ \Delta d(1) + \Delta \bar d(1) 
- 2[\Delta s(1) + \Delta \bar s(1)] = 0.58 \pm 0.15 \>. \nonumber 
\eea

Fig. 1 shows recent fits to $g_1$ data; details can be found
in Ref. \cite{vog}. The $Q^2$ QCD evolution is well reproduced at all
available $x$ values ; the lack of small $x$ data does not allow yet 
to constrain the gluon distribution, which only enters indirectly,
via density evolution. The dotted lines in the right plot of Fig. 1 
show variations on $g_1$ corresponding to the two possible extreme cases 
for $\Delta g$ at the initial scale: either $\Delta g = g$ or  
$\Delta g = - g$. The shaded 
areas correspond to a variation of $\Delta g(1, Q^2=5 \> ({\rm GeV}/c)^2)$
between $- 0.81$ and $1.73$. In these fits a $SU(3)_F$
symmetric polarized sea has been assumed, $\Delta u_s = \Delta \bar u
= \Delta d_s = \Delta \bar d = \Delta s = \Delta \bar s$; releasing
this assumption does not significantly change the quality of the fit. 

Inclusive polarized DIS scattering data do not allow yet a complete
precise determination of the gluon and the sea helicity distributions.

\subsection{Direct measurement of $\Delta g(x,Q^2)$}
 
Undoubtely, a direct measurement of $\Delta g(x, Q^2)$, in a process different
from inclusive DIS, is one of the most urgent and important issues in spin 
physics; there are many plans and projects for such a measurement, in 
running or planned experiments (HERMES, RHIC, COMPASS) or proposed ones 
(HERA-$\vec N$, TESLA-$\vec N$, ELFE, EIC, ...). A very first measure
of $\Delta g(x)/g(x)$ is already available, although with large errors and 
at one single value of $x$, from HERMES \cite{dg}.

One should look for $\Delta g$ in processes involving polarized gluons,
and measure spin asymmetries; one can isolate polarized gluon contributions
by carefully selecting final states and/or by selecting particular
kinematical regions where gluons are supposed to dominate.
Processes in which one could isolate $\gamma^* g$ elementary 
contributions are:

\begin{itemize}
\item
$\vec{\ell} \, \vec{N} \to \ell +$ 2 jets;
\item       
$\vec{\ell} \, \vec{N} \to \ell + h_1 + h_2 + X$, where $h_1$ and $h_2$ are 
large $p_T$ hadrons; 
\item
$\vec{\ell} \, \vec{N} \to \ell + c + \bar c + X$.
\end{itemize}
Quarks and gluons are supposed to initiate prompt photon production:
\begin{itemize}
\item
$\vec{p} \, \vec{N} \to \gamma X$,
\end{itemize} 
and $gg$ interactions might dominate in
\begin{itemize}
\item
$\vec{p} \, \vec{N} \to$ 2 jets $+ X$
\item
$\vec{p} \, \vec{N} \to h_1 + h_2 + X \>.$
\end{itemize}

\subsection{Flavour decomposition}

From data on inclusive DIS one only obtains information on 
combinations of $\Delta q + \Delta \bar q$, as Eq. (4) shows.
There is no direct way of separating $q$ and $\bar q$ contributions, 
there is no direct access to polarized sea distributions. 
This, instead, is possible in semi-inclusive DIS, $\ell N \to \ell h X$.

The double spin asymmetry for the $\gamma^* N \to h X$ process,
\be
A^h_1 \equiv \frac 
  {d\sigma^{\begin{array}{c}\hspace*{-0.1cm} \to \vspace*{-0.2cm}\\
\hspace*{-0.1cm}\Leftarrow\end{array}} 
-  d\sigma^{\begin{array}{c}\hspace*{-0.1cm} \to \vspace*{-0.2cm}\\ 
\hspace*{-0.1cm}\Rightarrow\end{array}}}
  {d\sigma^{\begin{array}{c}\hspace*{-0.1cm} \to \vspace*{-0.2cm}\\
\hspace*{-0.1cm}\Rightarrow\end{array}} 
+  d\sigma^{\begin{array}{c}\hspace*{-0.1cm} \to \vspace*{-0.2cm}\\
\hspace*{-0.1cm}\Leftarrow\end{array}}} \>,
\ee 
where the arrows now refer to the $\gamma^*$ ($\to$) and nucleon 
($\Rightarrow, \Leftarrow$) spin configurations, is given by
\be
A_1^h(x, Q^2) = \frac
{\sum_q e_q^2 \left[ \Delta q \, D_q^h + \Delta\bar q \, D_{\bar q}^h \right]}
{\sum_q e_q^2 \left[ q \, D_q^h + \bar q \, D_{\bar q}^h \right]}
\> [1 + R] \>.
\ee
$A_1^h$ is related by a simple kinematical factor to the measured asymmetry
$A^h_\parallel$, $R(x, Q^2)$ has the usual expression in terms of the 
unpolarized structure functions \cite{ael} and the $D_q^h(z,Q^2)$'s are the 
fragmentation functions, which now act as different weights for the
$\Delta q$ and $\Delta \bar q$ terms.  
   
Eq. (10) can be rewritten in terms of purities \cite{pur}, computable from
unpolarized distribution and fragmentation functions, which relate
measured quantities to polarized quark distributions. First data on 
$\Delta q_V \equiv \Delta q - \Delta \bar q$ and $\Delta \bar q$ are available
from HERMES \cite{dec} and much more are expected.  

\section{The polarized structure function \mbox{\boldmath $g_2(x, Q^2)$}}

New data on the polarized structure function $g_2$ have been recently 
published \cite{g2}, see Fig. 2, still with large errors and in limited
$x$ and $Q^2$ ranges. $g_2$ has higher-twist contributions, and no
direct partonic interpretation; it can be written as the sum of a
twist-2 contribution (the so called Wandzura-Wilczek piece \cite{ww}) and a 
higher-twist part:
\be
g_2(x, Q^2) = - g_1(x, Q^2) + \int_x^1 \frac{dy}{y} \> g_1(y, Q^2) 
+ g_2^{{\rm H-T}} \>.
\ee
A recent paper \cite{bkm} has studied the scale dependence of flavour
singlet contributions to $g_2$ and given an approximate expression 
of the higher-twist part in terms of quark and gluon correlations.
This expression satisfies the Burkhardt-Cottingham sum rule \cite{bc} 
$\int_0^1 g_2 \> dx = 0$.    

Data do not allow yet to extract the higher-twist contribution.
 
\begin{figure}[t]
\begin{center}
\epsfig{file=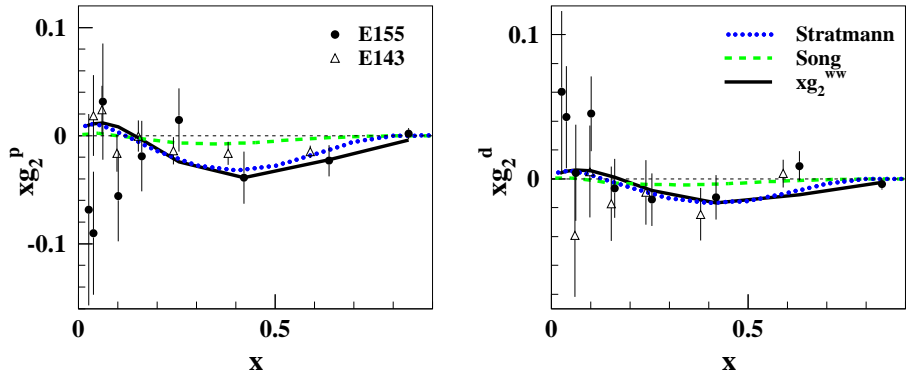,width=1.0\textwidth}
\caption{E155 data on $g_2$, and comparison with twist-2 Wandzura-Wilczek
contribution (solid line); see Ref. [5] for more details.}
\end{center}
\end{figure}

\section{Orbital angular momentum}

As partons in a nucleon are not collinear, according to angular
momentum conservation in the emission of a gluon by a massless quark,
the total helicity of a nucleon satisfies the spin sum rule
\be
\frac 12 = \frac 12 \Delta\Sigma(1,Q^2) + \Delta g(1,Q^2)
+ L_q(Q^2) + L_g(Q^2) 
\ee
where $\Delta \Sigma(1,Q^2) = \sum_q [\Delta q(1,Q^2) + \Delta \bar q(1,Q^2)]$
is twice the total helicity carried by quarks, $\Delta g(1,Q^2)$ is the total
helicity of gluons and $L_{q,g}(Q^2)$ is the total component of the 
orbital angular momentum of quarks, gluons along the motion direction. 

The issue of $L_q$ and $L_g$ is a subtle and controversial one; one should
wonder whether or not each of the terms in the above equation is gauge 
invariant, interaction independent, measurable and related to an integral
over a parton $x$ distribution. I urge the reader to consult the recent clear
discussions on this subject by Jaffe \cite{jaf}.

The outcome is that some proposed angular momentum operators~\cite{ji} for 
the different contributions to the nucleon spin, whose matrix elements in a
nucleon state might be measurable in deeply virtual Compton scattering, 
do not have a parton interpretation and are not gauge or interaction
independent; on the other hand, for some other definition of angular momentum 
operators~\cite{jaf} which do not suffer the same problems, there is no known 
way of measuring the corresponding matrix elements. 

From the analysis of Ref. \cite{vog} the orbital angular momentum 
contribution at the initial scale is $L_{q+g}(\mu^2) \simeq 0.15$.

\section{Skewed parton distribution} 

Having mentioned the deeply virtual Compton scattering, I have at least to 
comment on a very important issue, actually a whole new investigation region, 
which has also important spin aspects \cite{diehl}: that of the 
skewed or off-forward or generalized parton distributions. These are defined 
as the matrix elements between different nucleon states of the same operators
which, in the diagonal case, define the unpolarized and polarized 
partonic distributions. In different limits, infact, the skewed parton
distributions, give the partonic distributions or the elastic nucleon
form factors. 
 
Such distributions represent a novel field of investigation, offer new and 
deeper insights into the nucleon structure and detailed measurement programs
at JLAB, HERMES or at new dedicated machines have been discussed;
these off-forward matrix elements between polarized nucleon states generalize
helicity and transversity distributions. 
    
\section{Transversity}

The trasverse polarization of quarks inside a trasversely polarized nucleon,
denoted by $h_1$, $\delta q$ or $\Delta_T q$, is a fundamental twist-2 
quantity, as important as the unpolarized distributions $q$ and the 
helicity distributions $\Delta q$. It is given by
\be 
h_1(x, Q^2) = q_\uparrow^\uparrow(x, Q^2) - q_\downarrow^\uparrow(x,Q^2) \>,
\ee
that is the difference between the number density of quarks with transverse
spin parallel and antiparallel to the nucleon spin. It is the same as the
helicity distribution only in a non relativistic approximation,
but it is expected to differ from it for a relativistic nucleon.  

When represented in the helicity basis [see Fig. 3] $h_1$ relates quarks 
with different helicities, revealing its chiral-odd nature. This is the 
reason why this important quantity has never been measured in DIS:
the electromagnetic or QCD interactions are helicity conserving, there
is no perturbative way of flipping helicities and $h_1$ decouples from
inclusive DIS dynamics, as shown in Fig. 3a.

However, it can be accessed in semi-inclusive DIS, where some
non perturbative chiral-odd effects may take place in the non perturbative 
fragmentation process, Fig. 3b. Indeed, a serious program to measure $h_1$ 
in semi-inclusive DIS at HERMES, where a transversely polarized proton
target will soon be available, is in progress. We outline here two possible
ways of measurement~\cite{dan}.

\begin{figure}[t]
\begin{center}
\epsfig{file=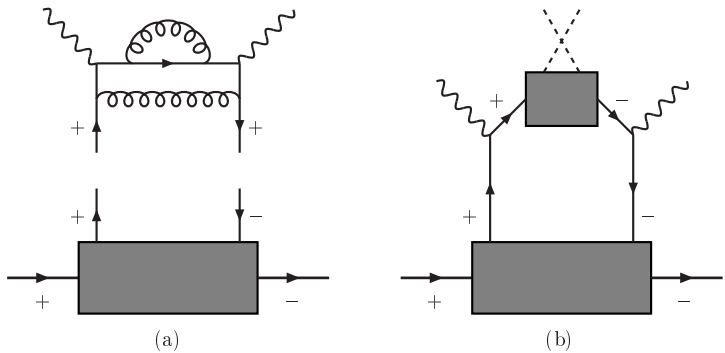,width=1.0\textwidth}
\caption{The chiral-odd function $h_1$ (lower box) cannot couple to inclusive
DIS dynamics, even with QCD corrections; it couples to semi-inclusive DIS,
where chiral-odd non perturbative fragmentation functions may appear.}
\end{center}
\end{figure}

\subsection{$h_1$ and the Collins function} 

The fragmentation process of a transversely polarized quark into a hadron
(say a pion) can have, according to Collins suggestion \cite{col}, a spin 
and intrinsic $\bfk_\perp$ dependence:
\be 
D^h_q(\bfp_q, \bfP_q; z, \bfk_\perp) = \hat D^h_q(z, k_\perp) + 
\frac 12 \> \Delta^ND^h_{\qup}(z, k_\perp) \> P_q \, \sin\Phi_C \>.
\ee
The quark with momentum $\bfp_q$ decays into a hadron with momentum
$\bfp_h = z \bfp_q + \bfk_\perp$ ($\bfk_\perp \cdot \bfp_q = 0$).  
$\bfP_q$ is the transverse quark polarization and the Collins angle 
$\Phi$ is the azimuthal angle between $\bfP_q$ and $\bfk_\perp$.
The Collins mechanism is depicted in Fig. 4. 

\begin{figure}[t]
\begin{center}
\epsfig{file=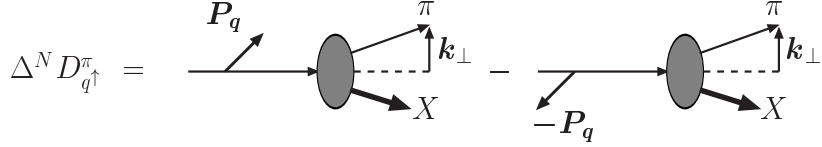,width=1.0\textwidth}
\caption{Pictorial representation of Collins function; notice that a similar
function is sometimes denoted by $H_1^\perp$ in the literature.}
\end{center}
\end{figure}

As a consequence, a single transverse spin asymmetry in the process 
$\ell \pup \to \ell \pi X$ can arise, at leading twist, from
the coupling of the chiral-odd transversity distribution $h_1$ with the 
chiral-odd Collins function $\Delta^ND^\pi_{\qup}$:
\be
A^h_N = \frac{\sum_q e_q^2 \, h_{1q}(x) \> \Delta^ND^h_{\qup}(z, k_\perp)}
{2\sum_q e_q^2 \, q(x) \> \hat D^h_q(z, k_\perp)} \>
\frac{2(1-y)} {1 + (1-y)^2} \> P \> \sin\Phi_C \>, \label{asym1} 
\ee
where $P$ is the magnitude of the transverse (with respect to the 
$\gamma^*$ direction) nucleon polarization.

Such an asymmetry has been observed for pions by HERMES~\cite{her} and 
SMC~\cite{smc} revealing that both $h_1$ and the Collins functions must be 
large \cite{am}; it promises to be -- apart from the importance of showing 
the Collins effect -- a viable access to the first measurement of $h_1$.

\subsection{$h_1$ and $\Lambda$ polarization}
Another way of accessing $h_1$ in semi-inclusive DIS is by measuring the 
transverse $\Lambda$ polarization in the process 
$\ell \pup \to \ell \Lup X$. At leading twist this is given by
\be
P_\perp = \frac
{\sum_q e_q^2 \, h_{1q}(x) \, \Delta_TD_q^\Lambda(z)}
{\sum_q e_q^2 \, q(x) \, D_q^\Lambda(z)}
\> \frac{2(1-y)}{1+(1-y)^2}
\ee
and it couples $h_1$ to the chiral-odd transversity fragmentation
$\Delta_T D_q^\Lambda = D_{\qup}^{\Lup} - D_{\qup}^{\Ldown}$. 
$\Lambda$ polarization is easily detectable via its angular decay
distribution.

\section{\mbox{\boldmath $\Lambda$} polarization in semi-inclusive DIS}

A measurement of $\Lambda$ polarization in polarized or unpolarized 
semi-inclusive DIS -- both with neutral and charged current contributions --
is very interesting in general and it allows to obtain new information on 
polarized distribution and fragmentation functions and to test at a 
subtle level the elementary dynamics. The following cases have been 
considered, with all possible spin configurations~\cite{lampol}:
\bea
\ell \, N  &\to& \ell \, \Lambda \, X \quad\quad\quad 
\nu  \, N  \to  \nu  \, \Lambda \, X \nonumber \\ 
\nu  \, N  &\to& \ell  \, \Lambda \, X \quad\quad\quad
\ell \, N  \to  \nu  \, \Lambda \, X \nonumber \>.
\eea

\section{Bloom-Gilman duality and elastic proton form factors}

I conclude by mentioning two recent interesting results from JLAB.
The first comes from the rich transition region from small to 
large $Q^2$ and it is a confirmation of the validity of the Bloom-Gilman 
duality -- the ``average'' value of the structure functions in the 
bumpy resonance region agrees with the large $Q^2$ behaviour --  
for the polarized structure function $g_1$ \cite{bg}.

The second concerns a measurement -- which unfortunately might be
the last one for a long time -- of the proton elastic form factors, defined 
in the $\gamma^*\,p$ elastic coupling by
\be
\Gamma^\alpha_{{\rm em}} = F_1(Q^2)\,\gamma^\alpha + \frac {\kappa}{2M} \>
F_2(Q^2) \, i \, \sigma^{\alpha \beta} q_\beta 
\ee
where $\kappa$ is the proton anomalous magnetic moment.
An old perturbative QCD prediction favoured the large $Q^2$ behaviours
\be
F_1 \sim \frac{1}{Q^4} \quad\quad\quad\quad F_2 \sim \frac{1}{Q^6} 
\ee
whereas the surprising result found at JLAB by the Hall A collaboration
\cite{jlab} shows that for $Q^2$ between 2 and 4 (GeV$/c)^2$
\be
Q \, F_2(Q^2) \sim F_1(Q^2) \>.
\ee
Preliminary results seem to confirm such a behaviour up to 
$Q^2 = 6$ (GeV/$c)^2$.

This result has immediately prompted some interesting theoretical 
considerations related to the role of quark transverse motion and orbital 
angular momentum \cite{ral}.

\section{Conclusions}  

I simply conclude by reminding some of the most recently 
obtained results and of the next ones expected, both theoretically and 
experimentally, in DIS spin physics.

\begin{itemize}
\item
Estimate of $\Delta g(x,Q^2)$ from QCD evolution in inclusive DIS and 
from direct measurements in other processes;
\item
measurement of $g_2(x,Q^2)$ and evaluation of its higher-twist component;
\item
theory of $h_1(x, Q^2)$ and its measurement;
\item
theory of quark and gluon orbital angular momentum and intrinsic $\bfk_\perp$;
\item
theory and measurement of off-forward parton distributions;  
\item
single azimuthal asymmetries in semi-inclusive DIS, new data with 
transversely polarized targets;
\item
$\Lambda$ polarization in polarized and unpolarized semi-inclusive DIS,
with neutral and charged currents; first measurement of polarized 
fragmentation functions; 
\item
flavour decomposition of spin distributions; 
\item
$\Delta q$ at very large and small $x$;
\item
elastic electromagnetic proton form factors;
\item
higher-twists, exclusive reactions, transition from low to large $Q^2$, ...

\end{itemize}
  
\section*{Acknowledgments}

I would like to thank the organizers for the invitation to deliver
this plenary talk and for the excellent organization.
I would like to thank U. D'Alesio and F. Murgia for help and suggestions 
in preparing the manuscript.


\section*{References}
\begin{thebibliography}{99}
\bibitem{emc}
J. Ashman {\it et al.}, \Journal{\PLB}{206}{364}{1988}.
\bibitem{al}
M. Anselmino and E. Leader, \Journal{\ZPC}{41}{239}{1988}.  
\bibitem{nf}
S. Forte, M.L. Mangano and G. Ridolfi, \Journal{\NPB}{602}{585}{2001};
M.L. Mangano {\it et al.}, Report of the nuDIS Working Group for the 
ECFA-CERN Neutrino- Factory study, CERN-TH/2001-131,
e-Print Archive: hep-ph/0105155.
\bibitem{ael}
See, {\it e.g.}, M. Anselmino A. Efremov and E. Leader, 
\Journal{\PR}{261}{1}{1995} 
\bibitem{g2}
E155 Collaboration, P.L. Anthony {\it et al.}, 
\Journal{\PLB}{458}{529}{1999}. 
\bibitem{fj}
For a recent review paper on the nucleon spin structure see
B.W. Filippone and X. Ji, e-Print Archive: hep-ph/0101224. 
\bibitem{dglap}
V.N. Gribov and L.N. Lipatov, {\em Sov. J. Nucl. Phys.} {\bf 15},
138 (1972); Y.L. Dokshitzer, {\em Sov. Phys. JETP} {\bf 16}, 161 (1977);
G. Altarelli and G. Parisi, \Journal{\NPB}{126}{298}{1977}.
\bibitem{vog}
For a recent analysis of polarized parton distributions see
M. Gl\"uck, E. Reya, M. Stratmann and W. Vogelsang,
\Journal{\PRD}{63}{094005}{2001}. 
\bibitem{dg}
HERMES Collaboration, A. Airapetian {\it et al.},
\Journal{\PRL}{84}{2584}{2000}. 
\bibitem{pur}
J.M. Niczyporuk and E.E.W. Bruins, \Journal{\PRD}{58}{091501}{1998}.
\bibitem{dec}
HERMES Collaboration, K. Ackerstaff {\it et al.}, 
\Journal{\PLB}{464}{123}{1999}.
\bibitem{ww}
S. Wandzura and and F. Wilczek, \Journal{\PLB}{72}{195}{1977}.
\bibitem{bkm}
V.M. Braun, G.P. Korchemsky and A.N. Manashov, \Journal{\NPB}{603}{69}{2001}.
\bibitem{bc}
H. Burkhardt and W.N. Cottingham, {\em Annals Phys.} {\bf 56}, 453 (1970). 
\bibitem{jaf}
R.L. Jaffe, e-Print Archive: hep-ph/0102281; e-Print Archive: hep-ph/0101280.  
\bibitem{ji}
X. Ji, \Journal{\PRL}{78}{610}{1997}.
\bibitem{diehl}
M. Diehl, talk in these Proceedings.
\bibitem{dan}
See also the talk by D. Boer in these Proceedings,
e-Print Archive: hep-ph/0106206   
\bibitem{col}
J.C. Collins, \Journal{\NPB}{396}{161}{1993}
\bibitem{her}
H. Avakian (on behalf of the HERMES collaboration), \Journal{\NPB}{79} 
{523}{1999} (Proc. Suppl.); HERMES Collaboration, A. Airapetian {\it et al.}, 
\Journal{\PRL}{84}{4047}{2000}; talks by D. Hasch and K. Oganessyan in 
these Proceedings. 
\bibitem{smc}
A. Bravar (on behalf of the SMC collaboration), 
\Journal{\NPB}{79}{520}{1999} (Proc. Suppl.).
\bibitem{am}
M. Anselmino and F. Murgia, \Journal{\PLB}{483}{74}{2000}.
\bibitem{lampol}
Talks by U. D'Alesio, O. Grebenyuk and F. Murgia in these Procedings; 
M. Anselmino, M. Boglione, U. D'Alesio and F. Murgia, 
e-Print Archive: hep-ph/0106055, to appear in {\em Eur. Phys. J.} {\bf C},
and references therein. 
\bibitem{bg}
Talk by W. Korsch in these Proceedings.
\bibitem{jlab}
Jefferson Lab Hall A Collaboration, 
M.K. Jones {it et al.}, \Journal{\PRL}{84}{1398}{2000}.   
\bibitem{ral}
Talk by J. Ralston in these Proceedings.
\end{thebibliography}
\end{document}